%% file: jetptloss.tex
\documentclass[aps,prd,
 ]{revtex4}
\usepackage{graphicx}
\usepackage{color}
\usepackage{slashed}
\usepackage{amsmath}
\input{macros.tex}

\newcommand\aas{{\scriptscriptstyle AA}}
\newcommand\alphas{\alpha_{\scriptscriptstyle S}}
\newcommand\antikt{anti-$k_{\scriptscriptstyle T}$}

\newcommand\nevt{N_{\rm evt\/}}
\newcommand\njet{N_{\rm jet\/}}
\newcommand\pt{p_{\scriptscriptstyle T}}
\newcommand\ppr{p_\perp}
\newcommand\qhat{\hat q}

\begin{document}

\title{Parton energy loss at LHC tests for a strongly coupled medium}
\author{Sourendu\ \surname{Gupta}}
\email{sgupta@theory.tifr.res.in}
\affiliation{Department of Theoretical Physics, Tata Institute of Fundamental
         Research,\\ Homi Bhabha Road, Mumbai 400005, India.}
\author{Rishi\ \surname{Sharma}}
\email{rishi@theory.tifr.res.in}
\affiliation{Department of Theoretical Physics, Tata Institute of Fundamental
         Research,\\ Homi Bhabha Road, Mumbai 400005, India.}
\begin{abstract}
 We construct a measure of transverse momentum loss of jets in nuclear
 collisions at LHC directly using measurements of jet cross sections
 in PbPb and pp collisions. The proposal is shown to be equivalent to
 $R_\aas$ and is equally straightforward to construct.  Using data from
 the ATLAS collaboration at two different collision energies, we show
 that the proposed measure has small statistical uncertainties. We argue
 that systematic errors can be easily improved over our estimates by the
 experimental collaboration to such an extent that it directly probes
 whether the jet-medium interaction is due to a strongly interacting
 medium or a weakly interacting plasma. We argue that the current data
 may marginally favour a strongly interacting medium. On the other hand,
 assuming that the medium is weakly interacting, we are able to provide
 estimates of the jet quenching parameter $\qhat$ which are in rough
 agreement with previously reported estimates.
\end{abstract}
\maketitle

Fast particles moving through matter lose energy and momentum by
bremsstrahlung. This principle is used in particle detectors. Results
from RHIC and LHC show unambiguously that jets produced in nuclear
(AA) collisions differ from their counterparts in proton-proton (pp)
collisions \cite{STAR:2005gfr}. This is interpreted as evidence that fast
particles also lose energy to matter through the strong interactions,
as was first hypothesized in \cite{Bjorken:1982tu, Gyulassy:1990ye,
Wang:1992qdg} and observed at RHIC through the suppression of hadrons
at high momenta \cite{STAR:2005gfr, PHENIX:2004vcz}.

At the LHC the reconstruction of jets is routine and results
for jet quenching were presented very soon after the initial runs
\cite{CMS:2011iwn}. In this case $R_\aas$ can be defined through the
ratio of differential jet cross sections,
\beq
R_\aas = \frac{d\sigma_\aas}{d\pt dy} \Bigg/ \frac{d\sigma_{pp}}{d\pt dy}
       \qquad{\rm where}\qquad \frac{d\sigma_\aas}{d\pt dy} =
     \frac1{\nevt T_\aas}\;\frac{d\njet}{d\pt dy},
\eeq{raa}
with $\njet$ being the number of jet events out of a total $\nevt$ events
in a given $\pt$ and $y$ bin for a fixed bin of centrality, and $T_\aas$
being the thickness function in PbPb collisions \cite{Miller:2007ri}.
In this paper we explore another representation of this difference. From
the same jet cross sections, we may obtain a transverse momentum shift,
$\Delta\pt$ defined by setting
\beq
\left.\frac{d\sigma_\aas}{d\pt dy}\right|_{\pt}
 = \left.\frac{d\sigma_{pp}}{d\pt dy}\right|_{\pt+\Delta\pt}
\eeq{shift}
As long as $R_\aas$ is less than unity over a large enough range of
$\pt$, $\Delta\pt$ must be positive in most of this range, since the
cross sections on both sides of the equation fall with increasing $\pt$.
$\Delta\pt$ is a direct measure of jet energy loss, the two being
linearly related.  The jet $\pt$ and $\Delta\pt$, averaged over the
rapidity acceptance and azimuthal angle, are easily computable fractions
of the jet energy $E$ and the energy loss $\Delta E$.  The relations
between them is easily incorporated into an experimentalist's jet Monte
Carlo. So we hope that the construction that we outline here is used in
future to report $\Delta\pt$, or even the Monte Carlo-derived quantity
$\Delta E$, as a direct output from the LHC experiments. Finally, we
note that this proposal for the construction of $\Delta\pt$ differs
from that proposed several times earlier, namely through the mismatch
in $\pt$ between multiple objects in the final state \cite{Gupta:1994mk,
Wang:1996pe, Brewer:2018dfs}, either multiple jets or a $\gamma$/Z and its
recoil jet. The earlier proposals have theoretically smaller systematic
uncertainties, but have significantly larger statistical uncertainties
because they use rarer events. Another approach which differs from ours
proceeds by parametrizing the jet spectra in terms of a few parameters
\cite{Spousta:2015fca, He:2018gks}.

All of the experiments at LHC report cross sections from fully
reconstructed jets.  The ATLAS experiment, for example, clusters
calorimeter tracks using the {\antikt} algorithm with different jet
opening angles $R$. For jet energy determination the subtraction
of the underlying event is done carefully including effects due
to flow. ATLAS reports a study of the change in $R_\aas$ with CM
energy using $\sqrt S=2760$ GeV \cite{ATLAS:2014ipv} and $\sqrt
S=5020$ GeV \cite{ATLAS:2018gwx}. ALICE \cite{ALICE:2019qyj} and CMS
\cite{CMS:2021vui} report no statistically significant variation of
$R_\aas$ over a large range of $R$.  Since all experiments report jet data
for $R=0.4$, we choose this jet definition for our analysis of the ATLAS
data. The results can then be cross validated by the other experiments.

At the lower energy ATLAS reports statistical errors of jet cross
sections in PbPb collisions to be 1.5--2.5\% over the energy range up
to $\pt=200$ GeV, growing to almost 10\% in the bin with $\pt>350$ GeV
in all centrality bins. The systematic uncertainties vary from about
15--20\%. Statistical errors for jets in pp collisions are smaller by a
factor of 3, and systematic uncertainties are around 3\%. At the higher
energy the systematic uncertainties in pp collisions decrease marginally
to about 2\%, but the statistical errors are currently larger, being
between about 10--15\%. At this energy, in PbPb collisions systematic
uncertainties are similar, and statistical errors range from 2--5\%
for $\pt\le200$ GeV but rise from 10--80\% at higher energy, depending
on the centrality. Clearly, the luminosity available at this energy
is insufficient to make a quantitative statement about jets in PbPb
collisions for $\pt>400$ GeV.

\bef
\begin{center}
\includegraphics[scale=0.75]{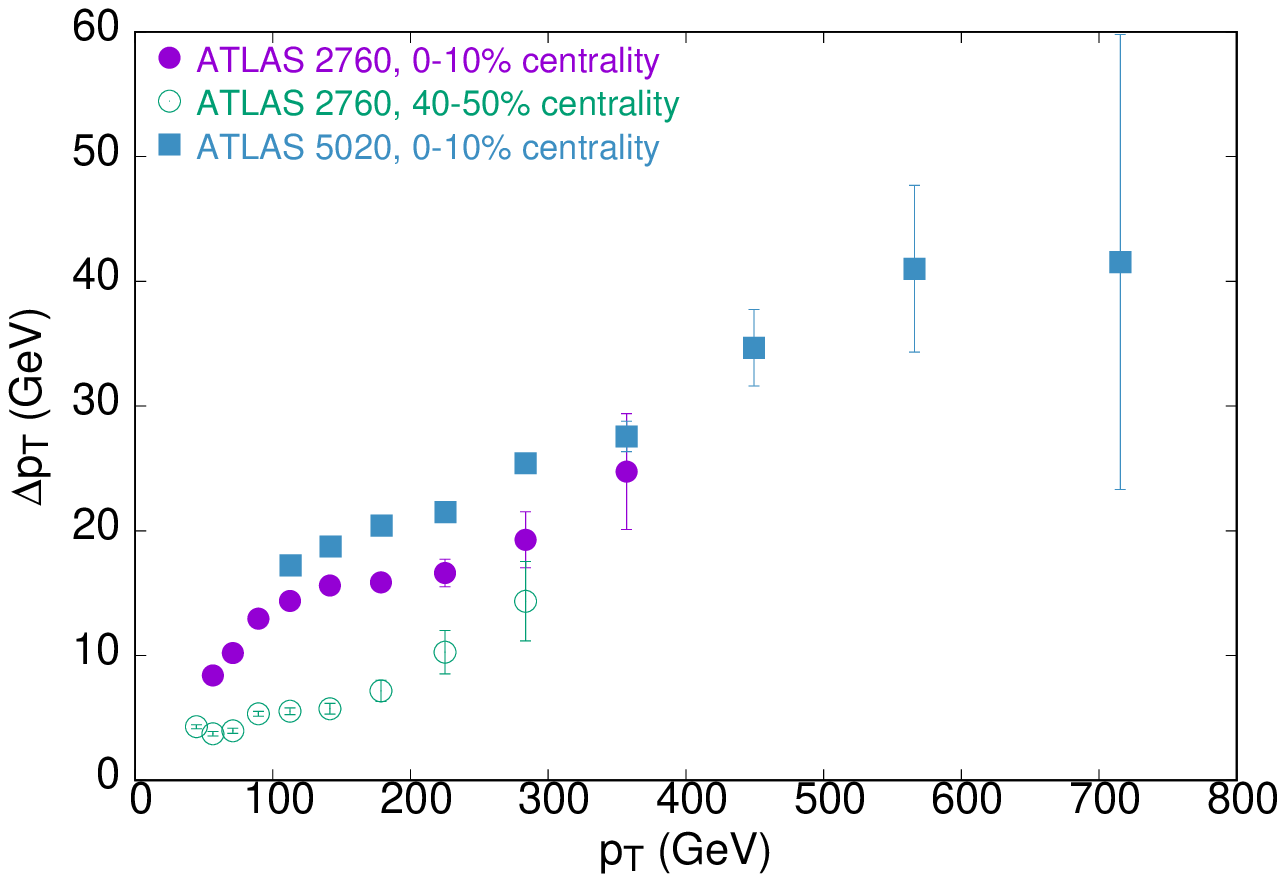}
\end{center}
\caption{$\Delta\pt$ inferred from the observations of the ATLAS
 experiment at $\sqrt S=2.76$ TeV \cite{ATLAS:2014ipv} in two centrality
 bins and at $\sqrt S=5.02$ TeV \cite{ATLAS:2018gwx}. The errors shown
 here come only from statistical errors.}
\eef{shift}

In \fgn{shift} we show the results for $\Delta\pt$ inferred from the cross
sections reported by the ATLAS experiment at some sample centrality bins
at two different collider energies. Similar results are obtained in all
centrality bins.  We discuss later how the shapes of these curves can
be used to distinguish between a strongly and a weakly coupled plasma,
and extract the jet energy loss parameters, $\qhat$, if the latter is
the case.  Errors in the jet cross sections in both PbPb and pp collisions
propagate into errors in determining $\Delta\pt$ through the definition
in \eqn{shift}. In the figure we show the error propagated into $\Delta\pt$
from the statistical errors in the cross sections.

The jet cross section in pp collisions shows a fairly steep fall
for $\pt<300$ GeV, and a slightly smaller logarithmic slope at higher
$\pt$. Unfortunately, this means that the large systematic uncertainty,
especially in PbPb collisions will translate into a large uncertainty
in $\Delta\pt$.  We may add these uncertainties in quadrature.
The justification is based on independent normal distributions of
errors. While this may be accurate for statistical errors, it is likely
to be far from correct for systematic uncertainties. Especially if there
are covariances between different sources, these need to be taken into
account. Since the correlation matrix between errors is not published,
it is hard for an independent analysis like ours to do full justice to
the error analysis.

One example of such a covariance in the systematic uncertainty which can be
corrected easily by the collaboration is that coming from uncertainties
in the luminosity. This affects the cross sections in both PbPb and pp
collisions, and therefore has a smaller effect on $\Delta\pt$. However,
since this is folded into the published tables of systematic errors,
in our analysis we have been forced to add them twice. There are similar
cancellations possible between some parts of the other systematic errors,
which can only be evaluated by the experimental collaboration.

We can try a crude estimation of the effect of covariances between different
components of the systematic uncertainties as follows. With the
$\Delta\pt$ determined through \eqn{shift} we can reconstruct $R_\aas$
through the definition
\beq
  R'_\aas(\pt) =
   \left.\frac{d\sigma_{pp}}{d\pt dy}\right|_{\pt+\Delta\pt}
   \Bigg/ \left.\frac{d\sigma_{pp}}{d\pt dy}\right|_{\pt}
\eeq{reconstruct}
This reconstructed quantity $R'_\aas$ is given its uncertainty band
through the usual error propagation algorithms. Comparing the propagated
error with the error in $R_\aas$ reported by the ATLAS collaboration gives
us a rough idea of the correction factor required from the covariance
of systematic uncertainties. In the left panel of \fgn{reconstruct} we
compare $R_\aas$ \cite{ATLAS:2018gwx} and $R'_\aas$. Two results follow
immediately. First, the estimates agree, showing that $\Delta\pt$ and
$R_\aas$ reported by the ATLAS collaboration are equivalent measures of
the difference between PbPb and pp jet cross sections.  Before making the
second inference, note that the bars shown here represent the statistical
and systematic uncertainties added in quadrature. For $R'_\aas$ the
uncertainty bars are reduced by a factor of five. As one can see from the
figure, this brings the two sets of uncertainties into rough agreement at
the highest $\pt$. This is an indication that systematic uncertainties
in $\Delta\pt$ that we can estimate using published data are an overestimate.
Covariances between systematic uncertainties are available to experimental
collaborations and can lead to a reduction of our naive computation of
uncertainties by a factor of five or more.

\bef
\begin{center}
\includegraphics[scale=0.55]{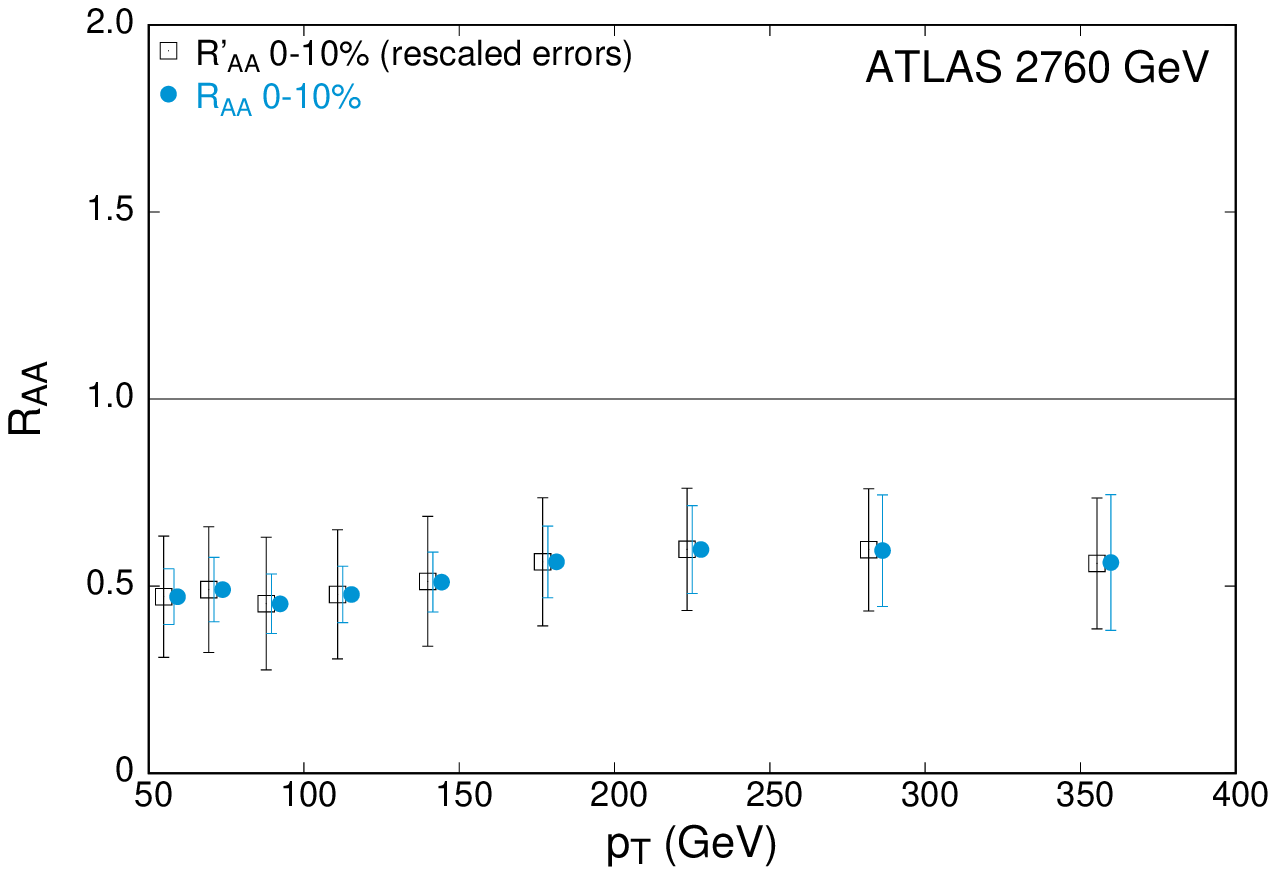}
\includegraphics[scale=0.55]{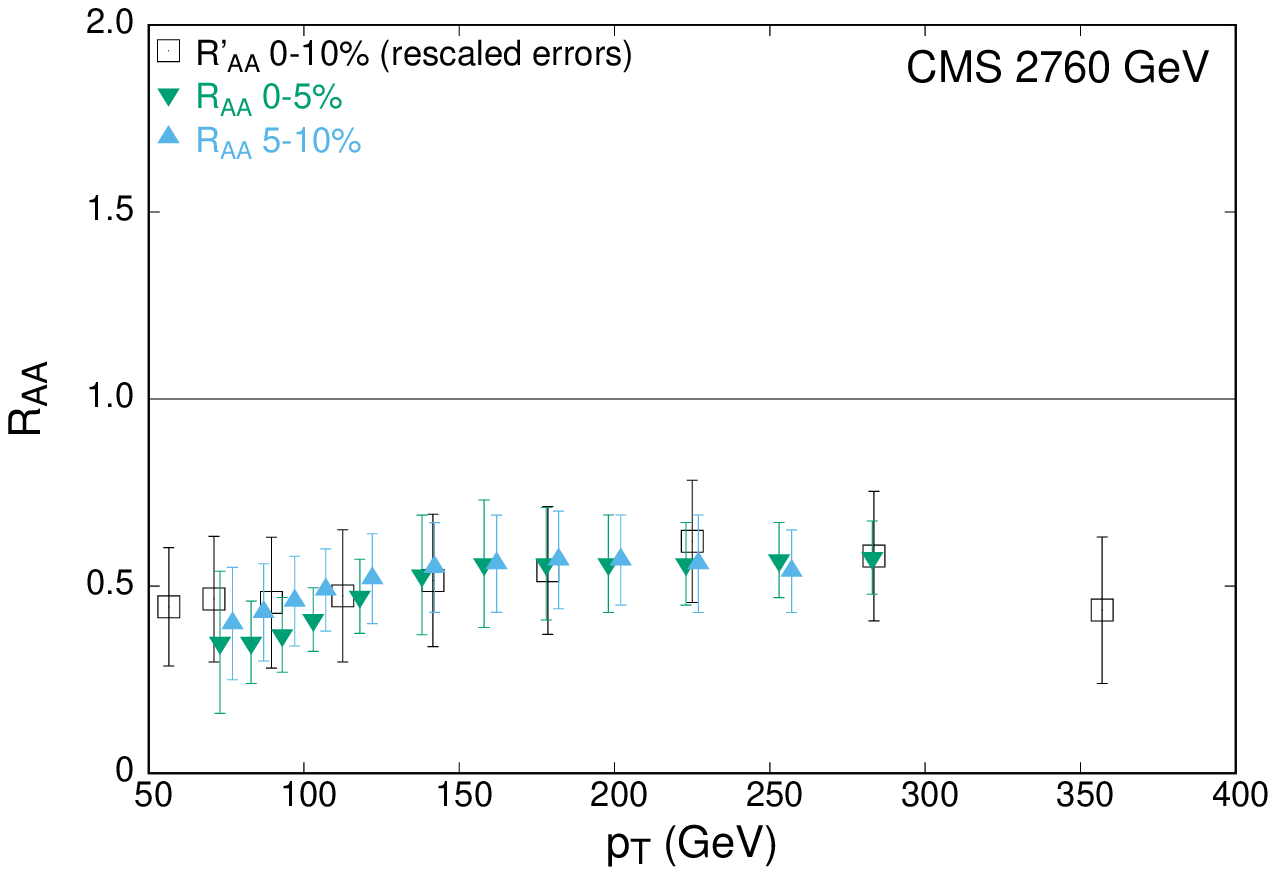}
\end{center}
\caption{Comparison of the reconstructed $R'_\aas$
 of \eqn{reconstruct} with direct experimental measurements of $R_\aas$
 from ATLAS (left) and CMS (right). The error bars shown are statistical
 and systematic uncertainties added in quadrature. The errors shown on
 $R'_\aas$ are one fifth of that obtained through error propagation on
 $\Delta\pt$ inferred from ATLAS observations at $\sqrt S=2.76$ TeV for
 the most central 0--10\% of events.  Experimental data are displaced
 slightly from the center of the $\pt$ bins for visibility.}
\eef{reconstruct}

The construction of $R'_\aas$ can also be used to test that $\Delta\pt$
from different experiments agree. For this we construct $R'_\aas$ using
the pp jet cross sections measured by the CMS \cite{CMS:2016uxf} along
with the $\Delta\pt$ that we extracted from the ATLAS measurements. The
latter are taken in the bin of 0--10\% centrality. The results agree
with $R_\aas$ reported by the CMS collaboration in the bins of 0--5\%
and 5--10\% centrality \cite{CMS:2016uxf}. A similar test is possible
for the data from the ALICE collaboration. However, the cross sections
reported are binned in the pseudo-rapidity $\eta$ instead of $y$
\cite{ALICE:2015mjv}, and with somewhat more stringent cuts on $\eta$. In
view of this, we do not show comparisons using these cross sections.

The results shown in \fgn{reconstruct} are a demonstration that
$R_\aas$ and $\Delta\pt$ are just two different ways of reporting
the experimentally observed differences between fully reconstructed
jets in pp and AA collisions. We claim that each representation has its
strengths. More than a decade of work has shown how useful $R_\aas$ is. In
the remainder of this letter we show that $\Delta\pt$ allows us to clearly
test very interesting physics, such as whether the fireball is strongly
or weakly coupled, and, in either case, answer more detailed questions
about the basic underlying theory. These demonstrations are made in order
to persuade experimental collaborations to extract and report $\Delta\pt$.

Jets are produced at the very earliest instants of the collisions,
and since they travel at the speed of light, $c$, they outstrip any
hydrodynamical disturbance, which can only travel at the speed of
sound, $v_s$. As a result, jets leave the fireball early, before a
rarefaction pulse can set fireball into collective transverse motion. So
they are a good primordial probe of matter in the fireball. A direct
experimental determination of the energy loss probes both thermal
and pre-thermal matter in the fireball, in principle. In some of the
analysis presented in this paper we will assume that soft effects from
the pre-equilibrium system are negligible, and $\Delta\pt$ is largely
due to the interaction of the jet with thermalized matter. Nevertheless
the ability of $\Delta\pt$ to probe pre-equilibrium physics must not
be forgotten. Possible tests of their importance could be to compare
experimental determinations with model results.

If the jet energy is $E$, then the 4-momentum of the parton, $P$, may
be written as $P=(E,0,0,E)$ when we choose the z-axis to be aligned
with the initial direction of motion of the hard parton in the final
state, and not the beam direction. Any interaction with thermal matter
(at temperature $T$) can push $P$ off-shell by adding momenta in a
perpendicular direction, $\ppr={\cal O}(T)$ changing the momentum
to $P+\ppr\simeq(E,T,0,E)$, for example.  The virtuality of the jet
is then $Q^2\simeq T^2$. However, any component of momentum added in
the longitudinal direction gives a contribution to the virtuality of
${\cal O}(ET)$. Since $E\gg T$, then it is more likely that $Q^2\simeq
ET$ \cite{Ovanesyan:2011xy}. When this virtuality is radiated away
by emitting a gluon, the coupling must be determined at the scale
$Q^2$. Even though the jet energy, $E$, is large, its interaction with
the medium is controlled by a strong coupling at a smaller scale. For
example, if $E\simeq100$ GeV, then the jet cross section is controlled
by $\alphas\simeq0.1$. However, if $T\simeq0.2$ GeV, then $Q\simeq5$
GeV, so that the jet-medium coupling is $\alphas\simeq0.2$--$0.3$
\cite{ParticleDataGroup:2020ssz}. The domain of $\alphas\le0.15$ is
definitely in the realm of weak coupling, and of $\alphas>0.5$ is often
considered to be strongly coupled.

In view of this a first question is whether the medium can be considered
as weakly coupled or not. If it is weakly coupled, then gluon radiation
is modeled as being created by a series of coherent collisions with
quasi-particles in the plasma \cite{Baier:1996kr, Zakharov:1996fv}
giving $\Delta\pt\propto L^2$, where $L$ is the path length of the jet
in the fireball (we mention a possible caveat later). On the other hand,
if the medium is strongly coupled, then it is modeled as exerting a
retarding force which causes gluon bremsstrahlung \cite{Marquet:2009eq}
giving $\Delta\pt\propto L^3$. Tests of these scalings have been attempted
before \cite{Horowitz:2007su, Jia:2011pi, Betz:2014cza}. We argue that
direct experimental access to $\Delta\pt$ allows us to make such and
more detailed tests, as we demonstrate.

Clearly, an important ingredient in all studies of the interaction of
hard probes with matter is the path length $L$. In a material whose
volume has a well-defined surface, $\Sigma$, this notion in simple. For
a jet produced at a point $\Pi$ inside $\Sigma$, with momentum pointing
in the direction $\hat{\bf P}$, walk along the ray $\Pi+s\hat{\bf
P}$. The distance $s$ at which the ray intersects $\Sigma$ is the path
length $L(\Pi,\hat{\bf P})$. The mean path length, $L$, is obtained
by averaging over $\Pi$ and $\hat{\bf P}$ for each impact parameter
$b$. Since a jet leaves the fireball before transverse expansion sets in,
one may examine the path length in a longitudinally expanding plasma
\cite{Gyulassy:2001kr}. This may be modelled as a boost-invariant
cylindrical region whose transverse shape is essentially given by
the initial collision geometry. Since matter interactions change the
jet-rapidity by an angle of order $T/E\ll1$, we may take its rapidity to
be fixed. Due to longitudinal boost invariance, it can then only leave
the fireball through the transverse surface $\Sigma$.

This simple model is complicated by two factors. One is that the
nuclear density is taken to be continuous in computing $T_\aas$, and
one has to take the same Wood-Saxon density \cite{DeVries:1987atn}
in this computation. The other is that during longitudinal expansion,
matter begins to diffuse outwards in the transverse direction. In
this case one has to estimate when the matter density is small enough
that the probability of a jet scattering with matter after travelling
distance $L$ is less than some pre-assigned value, $\epsilon$. Causal
diffusion equations are needed at such early times, and there are no
quantitative estimates of the two transport coefficients required
\cite{Gupta:2007bw}. In view of this uncertainty, we may place the
surface where 99\% of the matter is inside $\Sigma$. We find this radius
$R_{99} =1.53 R_{RMS}$, where the symbol on the right denotes the RMS
radius. This is, of course, a very generous over-estimate. Even in this
extreme scenario, we find that $L$ is 4.9 fm in the most central 0--10\%
of events in PbPb collisions, which is before transverse expansion is
well developed. Since $L$ decreases as one goes to more peripheral events,
jets can escape matter even earlier in these events.

\bef
\begin{center}
\includegraphics[scale=0.55]{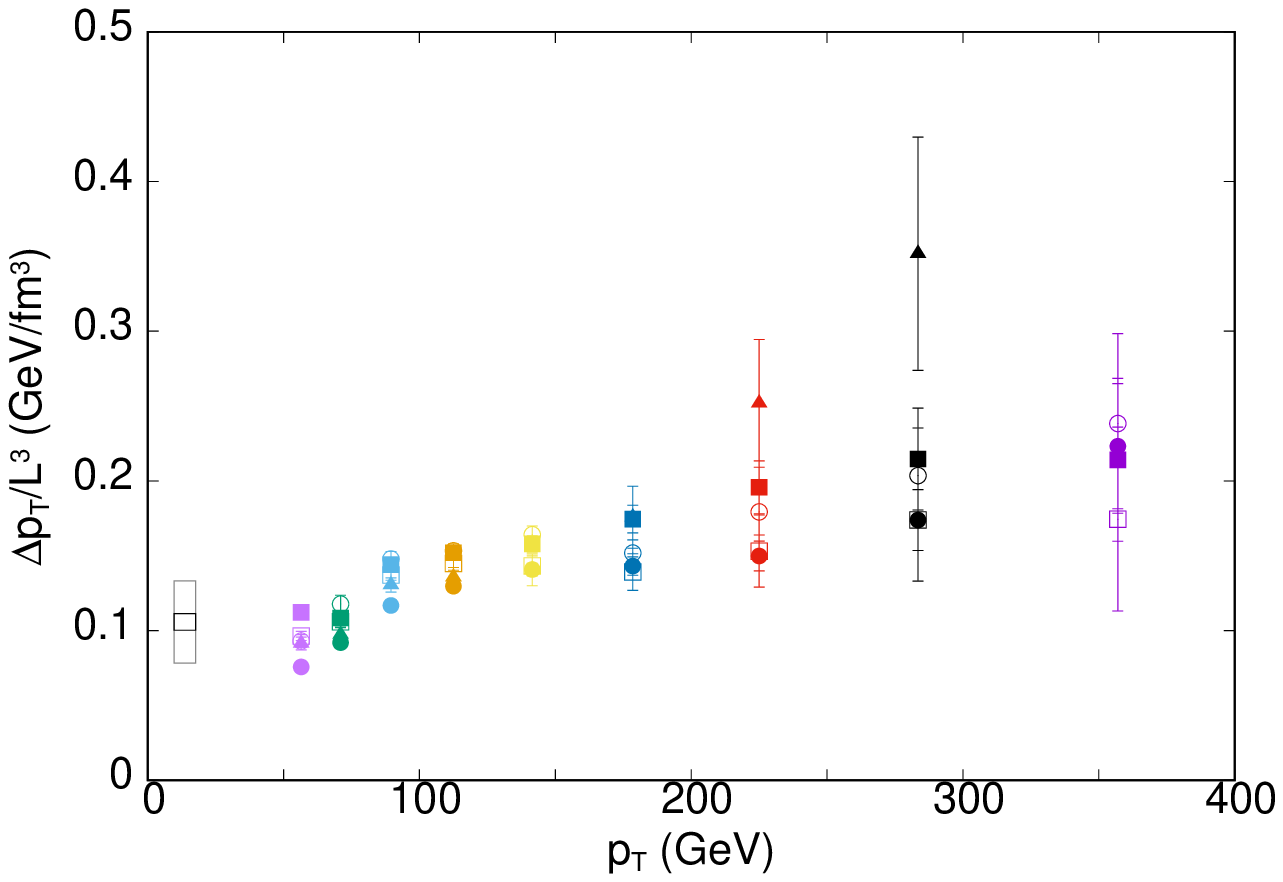}
\includegraphics[scale=0.55]{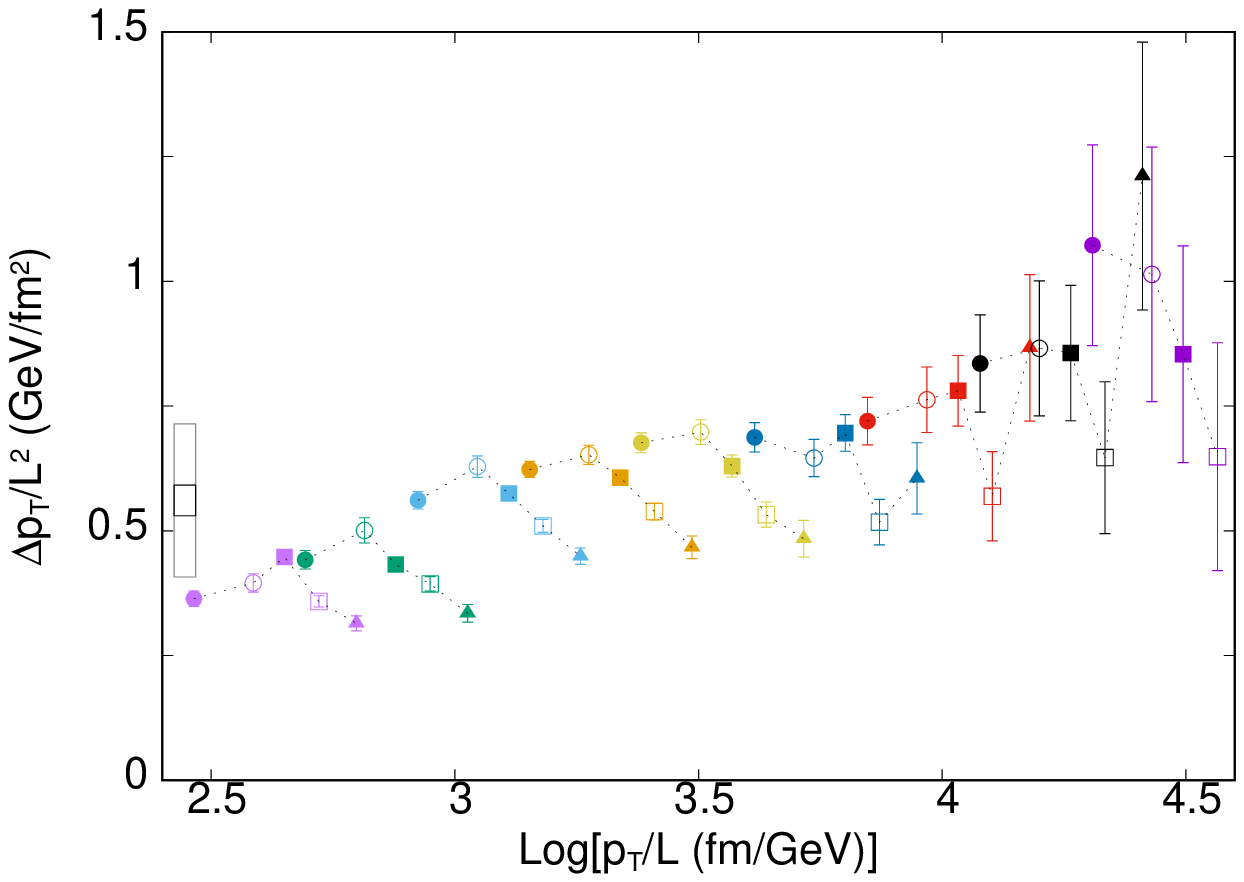}
\end{center}
\caption{Scaling with $L$ of the $\pt$ dependence of $\Delta\pt$
 at $\sqrt S=2.76$ TeV. The panel on the left tests for a strongly
 coupled plasma, that on the right for a weakly coupled medium. Perfect
 scaling would cause different centralities to collapse into an universal
 curve in $\pt$. In both plots results for the most central 0-10\%
 of events is denoted by filled circles, 10-20\% by unfilled circles,
 20--30\% by filled boxes, of 30--40\% by unfilled boxes, and 40-50\%
 by filled triangles. In the second panel, all the points in the same
 $\pt$ bin are joined by dashed lines.  The error bars are obtained
 using statistical errors on the cross sections. A typical magnitude
 of the uncertainty due to systematic errors is shown by the grey box
 at the extreme left of each panel, the darker inner box is the error
 after scaling by factor 5 as in \fgn{reconstruct}.}
\eef{scaling}

Although the simple hydrodynamical model that we have used is serviceable
enough, alternative initial conditions should be examined in future,
along with the hydrodynamic expansion of the fireball. We adopt this
simplified model for $L$ here because our treatment of experimental
systematic uncertainties currently dominate the errors in our treatment.
Once that is brought under better control, then one needs to improve
the computation of $L$, possibly by using a transport model.

In the remainder of our analysis, we suppose that non-radiative
collisional energy loss is a subdominant mechanism in the fireballs at
LHC. This has been tested in several publications \cite{Auvinen:2010yt,
Cao:2013ita}.  Alternative explanations of the observed jet energy loss,
namely through modification of parton densities and shadowing, have also
been quantified and found to be small \cite{CMS:2014qvs, Epple:2017qtk}.
If the radiative energy loss, \ie, bremsstrahlung, is the primary
mechanism of jet-matter interactions, then the centrality dependence
of $\Delta\pt$ must come from the dependence of $L$ on the impact
parameter $b$. This provides an easy test of whether or not matter is
strongly coupled. If it is, then $\Delta\pt/L^3$ should be independent
of centrality. On the other hand, if matter is weakly coupled, then
$\Delta\pt/L^2$ should be seen not to depend on centrality. We show these
tests in \fgn{scaling}. In the first panel one sees a test of whether the
medium is strongly coupled. At each $\pt$ we find that $\Delta\pt/L^3$
is independent of centrality with good accuracy within the statistical
uncertainties. The alternate hypothesis, of centrality independence of
$\Delta\pt/L^2$ is certainly ruled out if only statistical uncertainties
were taken into account. However, if one takes into account systematic
uncertainties, then it may seem that this hypothesis cannot be ruled
out. At $\sqrt S=5020$ GeV the statistical errors are larger and the
systematic errors are similar in magnitude. As a result, this test again
seems to be inconclusive unless correlations between systematic uncertainties
are taken into account.

Our previous discussion of the systematic uncertainties
in $R_\aas$ indicate that the situation can be improved by the LHC
experimental collaborations. Indeed, the simple prescription shown
in \fgn{reconstruct} which seems to tell us that covariances between
uncertainties can be roughly accounted for by reducing the uncertainties
by a factor of 5. In that case it is already clear that a test such as
that shown in \fgn{scaling} could disfavour the weakly coupled model
of the fireball. Although such a conclusion is currently premature, it
strongly indicates that if ATLAS, and other LHC collaborations, perform
the analysis shown, then they would easily be able to discriminate
between a strongly and weakly coupled medium.

\bef[b]
\begin{center}
  \includegraphics[scale=0.7]{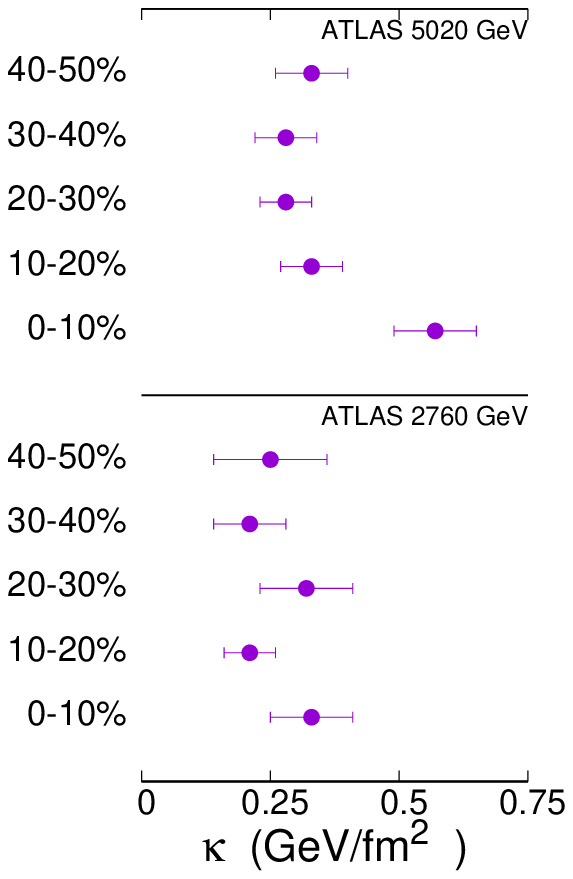}
  \includegraphics[scale=0.7]{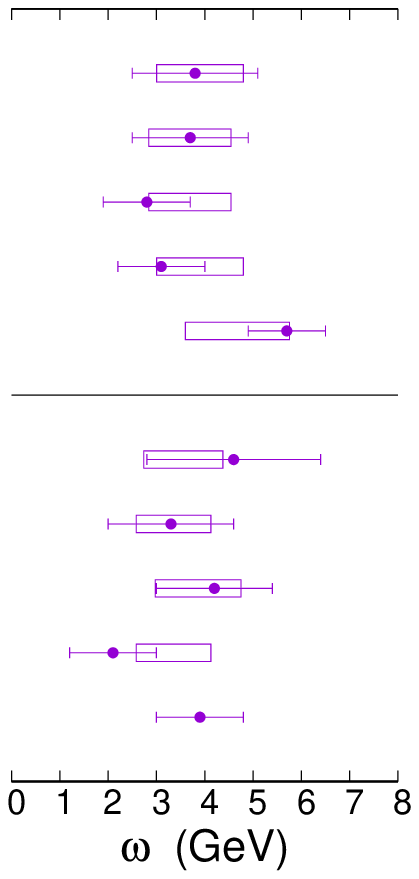}
\end{center}
\caption{The values of $\kappa$ and $\omega$ obtained by fits to data
 from \cite{ATLAS:2014ipv} and \cite{ATLAS:2018gwx}. The centrality
 and $\sqrt S$ for each set is indicated. Since the parameters are
 proportional to powers of $T$, the value of $\omega$ in any bin should
 be fixed by $\kappa$ in that bin and independent estimations of the two
 parameters in any one bin. The boxes in the panel on the right show a
 test of this hypothesis, with the uncertainty taken from the parameter
 uncertainty at the input 0--10\% centrality bin at $\sqrt S=2760$ GeV.}
\eef{weakparams}

Two further remarks are in order. First, that in strongly coupled
matter the ratio $\Delta\pt/L^3$ must be given by $\zeta T^4$. There
is no prediction of $\zeta$ for QCD, only for its conformal cousins
\cite{Liu:2006ug, Armesto:2006zv, Chernicoff:2012iq, Giataganas:2012zy}
(see however, \cite{Casalderrey-Solana:2014bpa} for an attempt to
parametrize the computation for QCD). It could depend on the 't-Hooft
coupling, which is formally $N_c\alphas(Q)$, and for QCD, \ie, with
$N_c=3$, may be of order unity. If the complete treatment of systematic
uncertainties do favour the strong coupling picture, then certainly the
sub-leading corrections in $N_c$ will have to be investigated. The figure
shows that over a large range of $\pt$ the result is compatible with
$T\simeq0.350$--0.4 GeV, if $\sqrt[4]{\zeta}$ is of order unity. The
minor variation seen in $\Delta\pt/L^3$ could come from the change in
$Q\simeq\sqrt{TE}$ and the logarithmic change it induces in $\alphas$.

On the other hand, if matter is weakly coupled, then $\Delta\pt/L^2$ is
clearly dependent on $\pt$. The treatment of BDMPS-Z \cite{Baier:2000mf}
shows that this is captured in the relation
\beq
 \Delta\pt = \kappa L^2\log\left(\frac{\pt}{\omega^2L}\right),
\eeq{bdmps}
where $\kappa\propto T^3$ and $\omega\propto T$. Further, one may write
$\kappa=C\alphas{\hat q}/4$ where $\alphas$ is the strong coupling at the
appropriate scale, and $C$ is given by the quadratic Casimir $C_F=4/3$
for quark initiated jets and $C_A=3$ for gluon initiated ones. For jets
at $y=0$, one finds that the gluon-gluon luminosity, $g^2(2\pt/\sqrt
S)$ dominates for $\pt<150$ GeV (for $\sqrt S=2.76$ TeV), beyond which
the quark gluon luminosity $g(2\pt/\sqrt S) \overline q(2\pt/\sqrt S)$
is larger. The jet cross sections due to these two parton subprocesses
are in the ratio $C_A/C_F=9/4$. The $gg$ subprocess contains $s$, $t$,
and $u$ channel exchanges, whereas the $\overline qg$ subprocess contains
only $s$ channel exchange. This causes another enhancement of the $gg$
subprocess by a factor of about 8. As a result, for jets with $\pt<400$
GeV, one may take the $gg$ subprocess to dominate. So one may take
$C=C_A$. Then taking $\alphas\simeq0.15$--0.25, as discussed earlier,
one has $\qhat=(5$--$9) \kappa$.

It is possible to extract the two parameters $\kappa$ and $\omega$ from
the values of $\Delta\pt$ at each centrality separately for each of the
two values of $\sqrt S$. In performing these fits we have added the
systematic and statistical uncertainties in quadrature. As discussed
earlier, this is an oversimplification, and direct access to the
results of detector Monte Carlos could allow a better treatment of the
systematic errors. The results are shown in \fgn{weakparams}.  There is
no statistically significant dependence of the parameters from one bin to
another. The minor variations are entirely accounted for by assuming that
there are small changes in $T$ as the centrality and $\sqrt S$ change.

We can extract an estimate for $\qhat$ given the range of $\kappa$
shown. This gives $\qhat=1$--4 GeV/fm${}^2$ (which is 0.2--0.8 GeV$^2$/fm
in units which have also been used in the literature). There are a
few previous estimates of $\qhat$ from jets at LHC. Among them we find
estimates ranging from $\qhat\simeq10\pm4$ GeV/fm$^2$ (at $T=470$ MeV
for $\sqrt S=2.76$ GeV) \cite{JET:2013cls}, $\qhat\simeq12.4$ GeV/fm$^2$
\cite{Mehtar-Tani:2021fud}, and $\qhat\simeq3$--7 GeV/fm$^2$ (for $T=470$
MeV) \cite{JETSCAPE:2021ehl}. This last range, in particular, is a 90\% CR
range, whereas we quote the more common 1-sigma errors, \ie, the 68\% CL.
Although our quoted range of $\qhat$ is on the lower side of the band, it
is consistent with the current spread of values, assuming that the medium
is weakly coupled. We note that if we had used $R_{RMS}$ to estimate $L$
we would have found $\qhat=2.3$--9.4 GeV/fm$^2$, which is in complete
agreement with the results of \cite{JET:2013cls, JETSCAPE:2021ehl}.

It is clear that more accurate results can be obtained not only through
better control of experimental systematic errors, but also by better
control of theoretical systematics on $L$. There is also the caveat that
we mentioned earlier. A possible correction to the formula in \eqn{bdmps}
would be to average $\kappa$ using the instantaneous temperature,
which changes as the system expands. If one assumed Bjorken expansion,
with $v_s=1/\sqrt3$, then one finds $\kappa$ should be replaced by
$\kappa_0\log(L/\tau_0)/ (L/\tau_0-1)$, where $\tau_0$ is the time at
which we can expect Bjorken expansion to set in and $\kappa_0$ is the
value of $\kappa$ at that time. When $L\gg\tau_0$ this could change the
$L$ dependence in \eqn{bdmps}. In particular, in this model, when $L$
is large one finds that the $L^2$ factor could change to $L\log L$. This
weaker dependence on $L$ would further reduce support for the weak
coupling picture, of course. However, this argument is approximate,
since a softer equation of state will make less of a difference, the
fuzzy borderline between pre-equilibrium and equilibrium expansion could
change the results, and there is always the question of how to separate
the scale of hydrodynamical expansion and the scale of coherence between
successive emissions that leads to the LPM effect. This last point is
clearly crucial since $TL$ is not much larger than unity, showing that
the scale separation is hard, and the assumption that $\qhat$ can be
replaced by its instantaneous value during hydrodynamic averaging could
be false. Clearly, answering the first question requires a hydro+transport
computation, whereas addressing this last requires a more careful analysis
of the transition between microscopic dynamics and bulk transport. These
interesting issues we leave to the future.

To summarize, we have introduced a measure $\Delta\pt$, the transverse
momentum loss of a hard jet in a medium, and shown that it can be
easily extracted from experimental measurements using \eqn{shift}. We
demonstrated that this although this information is exactly equivalent to
the widely used measure $R_\aas$, it can be used to directly distinguish
between weakly and strongly coupled plasmas. We have argued that the
presence of correlated systematic uncertainties makes it hard for people
outside the experimental LHC collaborations to get an accurate estimate
of errors in $\Delta\pt$.  However, very rough corrections for these
correlations/covariances shows that the data mildly supports a strongly
coupled fireball over one which is weakly coupled.  We have further shown
how one can extract medium properties from $\Delta\pt$.  In particular,
we showed that assuming that the fireball is weakly coupled, one can
extract a value of $\qhat$ which is consistent with current estimates.

We would like to thank Ankita Budhraja for very generously sharing the
complete results of her literature survey, in particular her compilation
of experimental results. We acknowledge support of the Department
of Atomic Energy, Government of India, under Project Identification
No. RTI 4002.

\end{document}

%% file: macros.tex
\newcommand\bef{\begin{figure}}
\newcommand\eef[1]{\label{fg:#1}\end{figure}}
\newcommand\beq{\begin{equation}}
\newcommand\eeq[1]{\label{#1}\end{equation}}
\newcommand\beqa{\begin{eqnarray}}
\newcommand\eeqa[1]{\label{#1}\end{eqnarray}}
\newcommand\bet{\begin{table}}
\newcommand\eet[1]{\label{tb:#1}\end{table}}

\newcommand\fgn[1]{Figure \ref{fg:#1}}
\newcommand\eqn[1]{eq.\ (\ref{#1})}

\newcommand\ie{{\sl i.e.\/}}